\documentclass{appolb}
\usepackage{graphicx}
\usepackage{amsmath}
\usepackage{authblk}

\begin{document}
\title{TOWARDS A NOVEL ENERGY DENSITY FUNCTIONAL FOR BEYOND-MEAN-FIELD CALCULATIONS WITH PAIRING AND DEFORMATION
\thanks{Presented at the Zakopane Conference on Nuclear Physics “Extremes of the Nuclear
Landscapes”, Zakopane, Poland, August 26 -- September 2, 2018}%
}
\author[1,2]{T. Haverinen}
\author[1,2]{M. Kortelainen} 
\author[1,3,4]{J. Dobaczewski}
\author[5]{K. Bennaceur}
\affil[1]{Helsinki Institute of Physics, \protect \\ P.O. Box 64, 00014 University of Helsinki, Finland}
\affil[2]{Department of Physics, University of Jyvaskyla \protect \\
P.O. Box 35 (YFL), 40014 University of Jyvaskyla, Finland} 
\affil[3]{Department of Physics, University of York, \protect \\ Heslington, York YO10 5DD, United Kingdom}
\affil[4]{Institute of Theoretical Physics, Faculty of Physics, University of Warsaw,
Pasteura 5, 02-093 Warszawa, Poland}
\affil[5]{IPNL, Universit\'e de Lyon, Universit\'e Lyon 1, \protect \\ CNRS/IN2P3, F-69622 Villeurbanne, France}
\maketitle
\begin{abstract}
We take an additional step towards the optimization of the novel finite-range pseudopotential at constrained Hartree-Fock-Bogolyubov level and implement an optimization procedure within an axial code using harmonic oscillator basis. We perform the optimization using three different numbers of the harmonic oscillator shells. We apply the new parameterizations in the O--Kr part of the nuclear chart and isotopic chain of Sn, and we compare the results with experimental values and those given by a parameterization obtained using a spherical code. 
\end{abstract}

\section{Introduction}
Novel approaches are essential when one aims to build an energy density functional (EDF) with spectroscopic quality and high predictive power, possibly applicable for beyond-mean-field calculations. 

The two most used families of non-relativistic nuclear EDFs are based on effective Skyrme and Gogny interactions. Despite their ability to reproduce nuclear binding energies fairly well, their shortcomings have also become apparent. The often used two-body density-dependent term, which is needed to reproduce some nuclear matter properties~\cite{(Dav18)}, introduces problems in beyond-mean-field and symmetry-restoration calculations -- for example, clear inconsistencies and anomalies can be seen in projected energies~\cite{(Dob07),(Ben09)}. Some strategies have been implemented to handle the problem with singularities, but there is no general solution for these problems unless the total energy is directly derived as an expectation value of a true interaction that is called functional generator~\cite{(Rai14),(Dob16)}.

Concerning the predictive power, recent analyses point out to the fact that the uncertainties of state-of-the-art models increase rapidly when going towards both the proton and neutron rich nuclei (see Ref.~\cite{(Hav17)} and references cited therein). In addition, these models miss some important physics, since differences between theoretical calculations and experimental results cannot be explained by statistical errors. 

Thus, to achieve significant improvements, novel approaches are called for. One possible direction is an EDF generated by a finite-range pseudopotential~\cite{(Dob12), (Ben17)}. The first EDF parameter adjustment gave promising results~\cite{(Ben17)}. However, propagated errors in deformed nuclei were found to be large, emphasizing the need for input data to constrain deformation properties. Furthermore, if the adjusted parameters are meant to be used for deformed nuclei with a code using a harmonic oscillator (HO) basis, it is interesting to study the dependence of parameters and statistical errors on the dimension of the basis.

\section{Methods}
We follow the definitions of the finite-range pseudopotential introduced in previous studies~\cite{(Dob12), (Ben17)}.
The different orders \(n\) of the pseudopotential are written as
\begin{align*}
\mathcal{V}_j^{(n)} \left( \mathbf{r_1}, \mathbf{r_2};\mathbf{r_3}, \mathbf{r_4}\right) = & \left( W_j^{(n)} \hat{1}_{\sigma} \hat{1}_{\tau} + B_j^{(n)} \hat{1}_{\tau} \hat{P}^{\sigma} - H_j^{(n)} \hat{1}_{\sigma} \hat{P}^{\tau} - M_j^{(n)} \hat{P}^{\sigma} \hat{P}^{\tau} \right) \\ & \times \hat{O}_j^{(n)} \left( \mathbf{k_{12}}, \mathbf{k_{34}} \right) \delta(\mathbf{r_{13}}) \delta(\mathbf{r_{24}}) g_a(\mathbf{r_{12}}),
\end{align*}
where \(\mathbf{k_{ij}} = \frac{1}{2i} (\mathbf{\nabla_i}- \mathbf{\nabla_j})\) is the relative momentum operator, \(\mathbf{r}_{ij}=\mathbf{r}_i-\mathbf{r}_j\) is the relative position, and \(\hat{P}^{\sigma}\) (\(\hat{P}^{\tau}\)) is the spin (isospin) exchange operator. 
We used a Gaussian form for the regulator
$g_a(\mathbf{r}) = \frac{1}{(a\sqrt{\pi})^3} \mathrm{e}^{-\mathbf{r}^2/a^2}$
with $a=1.15$~fm. There are three operators \(\hat{O}_j^{(n)}\) up to next-to-leading order (NLO), namely
\( \hat{O}_1^0(\mathbf{k_{12}},\mathbf{k_{34}}) = \hat{1} \), 
\(\hat{O}_1^1(\mathbf{k_{12}},\mathbf{k_{34}}) = \frac{1}{2} (\mathbf{k_{12}^{*2}} + \mathbf{k_{34}^{2}}) \) and
\(\hat{O}_2^1(\mathbf{k_{12}, \mathbf{k_{34}}}) = \mathbf{k_{12}^{*} \cdot \mathbf{k_{34}}}\). 
In addition, the zero-range Skyrme
\begin{align}
\mathcal{V}_{\delta} (\mathbf{r_1,r_2;r_3,r_4}) = t_0 \left( 1+x_0 \hat{P}^{\sigma}\right) \delta(\mathbf{r_{13}}) \delta (\mathbf{r_{24}}) \delta(\mathbf{r_{12}}),
\end{align}
was used with  \(x_0=1\) and \(t_0=1000 \) MeV \(\mathrm{fm}^3\). This term is active in the particle-hole channel only and counteracts the strong attraction from the finite-range term needed to obtain pairing strong enough in the bulk.
We considered only the local version of pseudopotential defined by the condition \(\hat{O}_i (\mathbf{k_{34}}+\mathbf{k_{12}}) = \hat{O}_i(\mathbf{k_{34}}-\mathbf{k_{12}^{*}})\), and as a consequence, parameters were coupled so that \(W_2^1 = -W_1^1\), \(B_2^1 = -B_1^1\), \(H_2^1 = -H_1^1\) and \(M_2^1 = -M_1^1\). In the end, there were 9 parameters to be optimized at local NLO, since one constant defining the zero-range spin-orbit term was also optimized.

We optimized parameters \(\mathbf{p}\) of the functional by minimizing the penalty function,
\begin{equation}
\chi^2(\mathbf{p}) = \sum_{i=0}^{N_d} \frac{(\mathcal{O}_i(\mathbf{p})-\mathcal{O}_i^{exp})^2}{\Delta \mathcal{O}_i^2},
\end{equation} 
where \(N_d\) represents the number of data points, \(\mathcal{O}_i(\mathbf{p})\) and \(\mathcal{O}_i^{exp}\) correspond to theoretical and experimental values of chosen (pseudo-)observables, respectively, and \(\Delta \mathcal{O}_i\) represents the tolerance related to the specific data point. 
Since the purpose of this study was to quantify the effects of the used model space size and compare the results with a parameterization optimized in a coordinate space, the data set followed the one of Ref.~\cite{(Ben17)}. The only exception was the average neutron pairing gap \(\langle \Delta_n \rangle\) in \({}^{120}\)Sn. In Ref.~\cite{(Ben17)} the average neutron gap was calculated with \(\ell_{max} = 9\) and \(\ell_{max} = 11\) (see~\cite{(Ben17)} for the definition of \(\ell_{max}\)). In this study it was calculated only once with the same set-up as the other data points. Otherwise the data set consisted of binding energies and radii of 8 doubly magic and semi-magic nuclei, namely \({}^{40}\)Ca, \({}^{48}\)Ca, \({}^{56}\)Ni, \({}^{78}\)Ni, \({}^{100}\)Sn, \({}^{120}\)Sn, \({}^{132}\)Sn and \({}^{208}\)Pb, and altogether six data points of pseudo-observables in the infinite and polarized nuclear matter. Finally, to avoid isovector finite-size instabilities, we used the isovector density \(\rho_1 (\mathbf{r})\) in the center of \({}^{208}\)Pb, and we aimed for the value \(\rho_1(0) > 0\), in the very same manner as described in Ref.~\cite{(Ben17)}.

We used the axial code HFBTEMP~\cite{(KorTBP)} together with the optimization algorithm POUNDerS~\cite{(Wil15)} for derivative-free nonlinear least squares problems. The code HFBTEMP expands the solutions of the HFB equations on the axial HO basis and we will use it for fits including deformed nuclei in future. We benchmarked HFBTEMP successfully against HFODD~\cite{(Sch17)}, and POUNDerS was already applied earlier in the field of nuclear physics~\cite{(Kor12)}. We optimized the local NLO pseudopotential by using three different numbers of the HO shells, namely 10, 12, and 14, whereby we obtained three parameterizations that we discuss. 


\section{Results}

\begin{figure}[htb]
\centerline{%
\includegraphics[width=9cm]{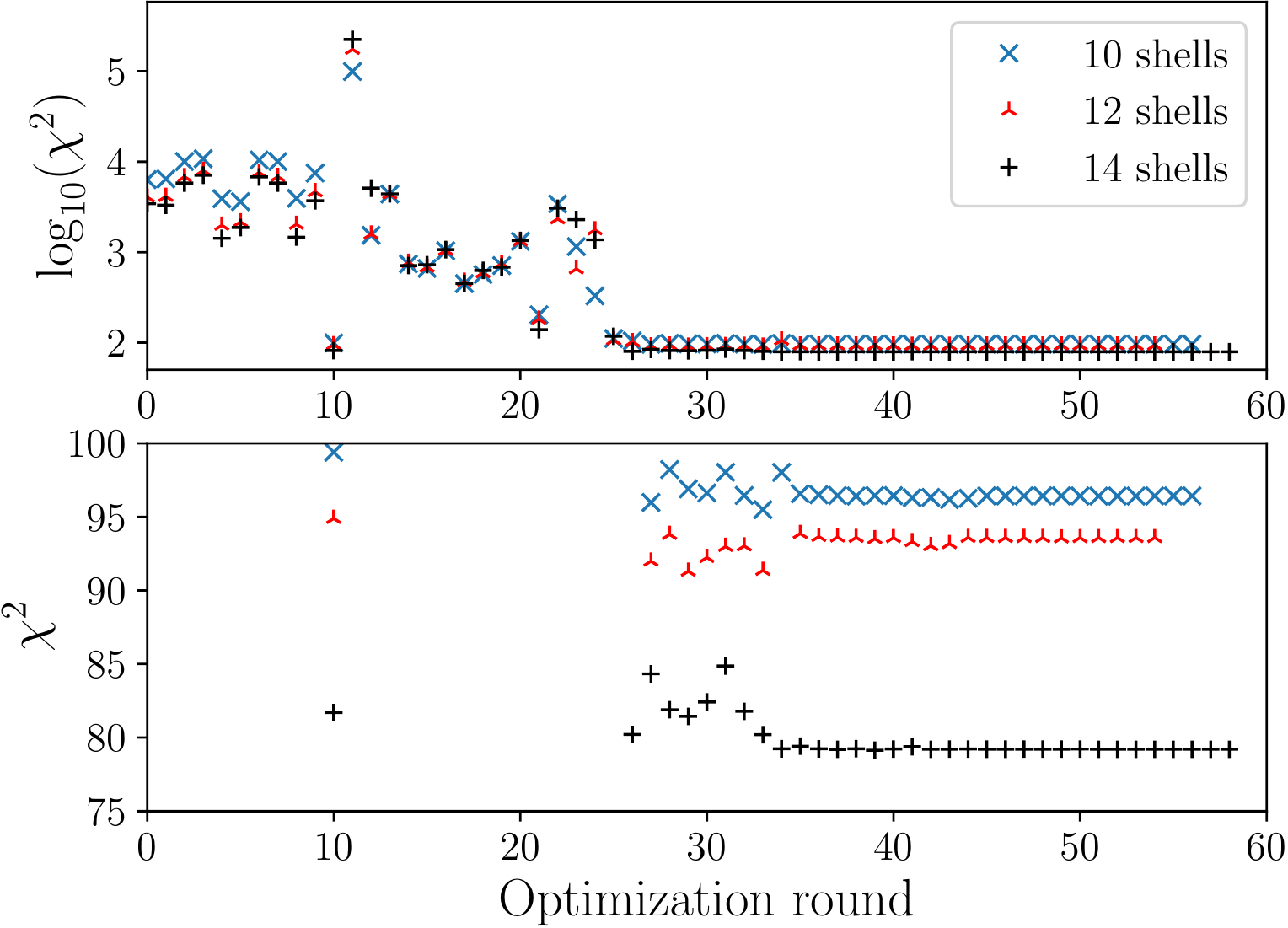}}
\caption{The objective function \(\chi^2\) as a function of the optimization rounds.}
\label{fig:chis}
\end{figure}

The convergence of the optimization procedure is shown in Fig.~\ref{fig:chis}. In the upper panel, the values of the objective function \(\chi^2\) are represented in a logarithmic scale as a function of the number of the optimization round, whereas in the lower panel the scale of \(\chi^2\) is natural. We observe that the required number of optimization iterations does not significantly depend on the used number of the HO shells. However, the needed computational time for every iteration is, of course, greater when a larger model space is used.

\begin{figure}[htb]
\centerline{%
\includegraphics[width=11cm]{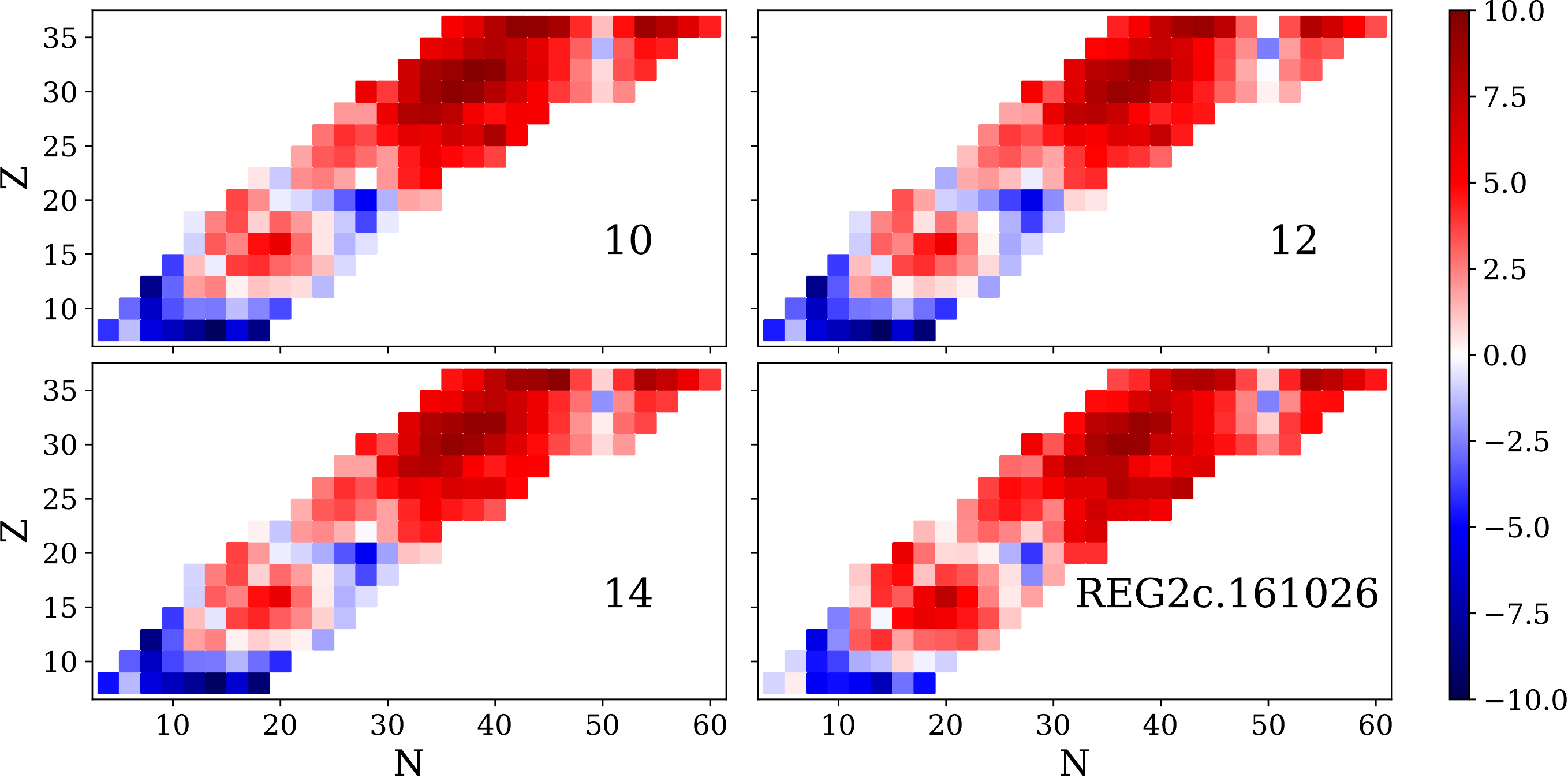}}
\caption{Binding energy residuals \(E_{Th}-E_{Exp}\) in units of MeV, calculated with parameter set optimized for 10, 12, and 14 HO shell basis. These are compared to results with REG2c.161026 parameterization~\cite{(Ben17)}, calculated here with 14 HO shells.}
\label{fig:charts}
\end{figure}

Using the same set-up as in the optimization, we tested the three obtained parameterizations by calculating even-even nuclei in the O--Kr part of the nuclear chart. The obtained binding energies are shown in the form of residuals \(E_{Th}-E_{Exp}\) in Fig.~\ref{fig:charts}. Here, all calculations were done at axially deformed HFB level, that is, for each nucleus we obtained the energy minimum with respect to deformation. Experimental binding energies were taken from AME2016 atomic mass evaluation~\cite{(Wan17)}. We compare the residuals to the ones given by the parameterization REG2c.161026 of Ref.~\cite{(Ben17)}, that was obtained by using the spherical coordinate space code \($FINRES$_4\)~\cite{(BenTBP)}. In this study, the theoretical binding energies given by REG2c.161026 were computed with HFBTEMP and 14 HO shells, assuming axial symmetry.  

\begin{figure}[htb]
\centerline{%
\includegraphics[width=10cm]{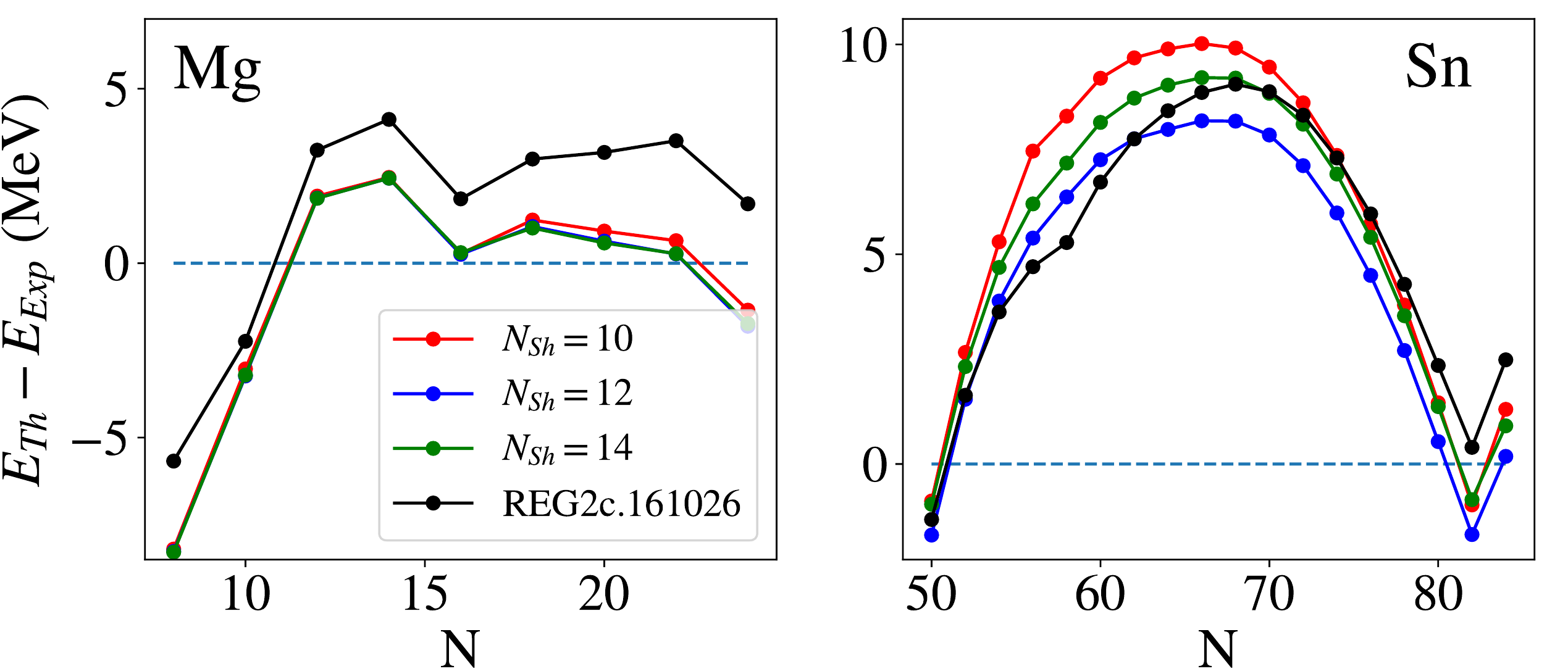}}
\caption{Same as in Fig.~\ref{fig:charts}, but for the isotopic chains of Mg and Sn and the residual \(E_{Th}-E_{Exp}\) represented on the ordinate.}
\label{fig:Mg_Sn_res}
\end{figure}

We observe that the differences between the results obtained with the parameters adjusted for 10, 12, and 14 HO shells are small in this part of the nuclear chart. These results differ more from the ones obtained by REG2c.161026, but still the differences are minor in mid-shell nuclei when comparing to the values of residuals. This can be seen also in Fig.~\ref{fig:Mg_Sn_res}, which represents the binding energy residuals of Mg and Sn nuclei as functions of the neutron number. Fig.~\ref{fig:Mg_Sn_res} shows how the binding energy residuals are greater in mid-shell nuclei, as expected, and how the binding energy residuals are not necessarily smaller if a larger model space is used. Our results for REG2c.161026 parameterization give less bound light nuclei since it was optimized with a code using coordinate space representation. This effect fades away in heavier nuclei, since a smaller basis state set can no longer accommodate all relevant aspects of a coordinate-space-based HFB solution.

\section{Summary and outlook}
We optimized a local finite-range pseudopotential up to next-to-leading order by using 10, 12, and 14 HO shells. We applied the three parameterizations in the O---Kr part of the nuclear chart and Sn isotopic chain. The obtained differences of computed binding energies turn out to be relatively small. This reflects the fact that even though the binding energies do depend on the number of used HO shells, this dependence is fairly well absorbed in the parameters during the optimization. Nevertheless, importance of the larger model space increases in heavier nuclei. The next step in the optimization of the pseudopotential will be to include data on deformed
nuclei in the penalty function, and the work in this direction is in progress. \\

T.H. was supported by Finnish Cultural Foundation, North Karelia Regional Fund (grant 55161255). J.D. was supported by the STFC grants Nos.~ST/M006433/1 and ST/P003885/1. We acknowledge the CSC-IT Center for Science Ltd., Finland, for the allocation of computational resources. 


\begin{thebibliography}{10}

\bibitem{(Dav18)}
D.~Davesne {\em et al.}
\newblock {\em Phys. Rev. C} 97, 044304 (2018).

\bibitem{(Dob07)}
J.~Dobaczewski {\em et al.}
\newblock {\em Phys. Rev. C} 76, 054315 (2007).

\bibitem{(Ben09)}
M.~Bender, T.~Duguet, D.~Lacroix.
\newblock {\em Phys. Rev. C} 79, 044319 (2009).

\bibitem{(Rai14)}
F.~Raimondi, K.~Bennaceur, J.~Dobaczewski.
\newblock {\em J. Phys. G: Nucl. Part. Phys.} 41, 055112 (2014).

\bibitem{(Dob16)}
J.~Dobaczewski.
\newblock {\em J. Phys. G: Nucl. Part. Phys.}
  43, 04LT01 (2016).

\bibitem{(Hav17)}
T.~Haverinen, M.~Kortelainen.
\newblock {\em J. Phys. G: Nucl. Part. Phys.}
  44, 044008 (2017).

\bibitem{(Car08)}
B.~G. Carlsson, J.~Dobaczewski, M.~Kortelainen.
\newblock {\em Phys. Rev. C} 78, 044326 (2008).

\bibitem{(Rai11)}
F.~Raimondi, B.~G. Carlsson, J.~Dobaczewski.
\newblock {\em Phys. Rev. C} 83, 054311 (2011).

\bibitem{(Dob12)}
J.~Dobaczewski, K.~Bennaceur, F.~Raimondi.
\newblock {\em J. Phys. G: Nucl. Part. Phys.}
  39, 125103 (2012).

\bibitem{(Ben17)}
K.~Bennaceur {\em et al.}
\newblock {\em J. Phys. G: Nucl. Part. Phys.}
  44, 045106 (2017).

\bibitem{(KorTBP)}
M.~Kortelainen.
\newblock {HFBTEMP} code,
\newblock {\em unpublished}.

\bibitem{(Wil15)}
S.~M. Wild, J.~Sarich, N.~Schunck.
\newblock {\em J. Phys. G: Nucl. Part. Phys.}
  42, 034031 (2015).

\bibitem{(Sch17)}
N.~Schunck {\em et al.}
\newblock {\em Comput. Phys. Commun.} 216, 145--174 (2017). 

\bibitem{(Kor12)}
M.~Kortelainen {\em et al.}
\newblock {\em Phys. Rev. C} 85, 024304 (2012).

\bibitem{(Wan17)}
M.~Wang {\em et al.}
\newblock {\em Chin. Phys. C} 41, 030003 (2017).

\bibitem{(BenTBP)}
K.~Bennaceur.
\newblock {FINRES}${}_4$ code,
\newblock {\em to be published}.

\end{thebibliography}

\end{document}